\newcommand{\CLASSINPUTinnersidemargin}{18mm}
\newcommand{\CLASSINPUToutersidemargin}{12mm}
\newcommand{\CLASSINPUTtoptextmargin}{20mm}
\newcommand{\CLASSINPUTbottomtextmargin}{25mm}
\documentclass[10pt,conference,a4paper]{IEEEtran}

\usepackage{times}

\usepackage{graphicx}

\usepackage{xcolor}
\usepackage{amsmath}
\usepackage{times}
\usepackage{graphicx}
\usepackage{amsfonts}
\usepackage{amssymb}
\usepackage{cite}
\usepackage{subfigure}
\usepackage{url}
\usepackage{float}
\usepackage{booktabs}
\usepackage{fancyhdr}

\DeclareGraphicsExtensions{.png,.eps,.ps,.pdf}

\begin{document}

\fancypagestyle{firstpage}
{
    \fancyhead[L]{Accepted at URSI 2021, Vigo}    
    \fancyhead[R]{20 - 24 September, 2021}
}

\title{Sound Event Localization and Detection Using Squeeze-Excitation Residual CNNs}


\author{
\authorblockN{Javier Naranjo-Alcazar$^{1,2}$,
      Sergi Perez-Castanos$^{1}$,
      Pedro Zuccarello$^{1}$, 
      Maximo Cobos$^{2}$,
      Francesc J.Ferri$^{2}$}
\authorblockA{\{javier.naranjo, sergi.perez, pedro.zuccarello\}@visualfy.com, \{maximo.cobos, francesc.ferri\}@uv.es}
\authorblockA{$^{(1)}$ Visualfy, Benisan\'o, Val\`encia, Spain}
\authorblockA{$^{(2)}$Departament d'Inform\`atica, Universitat de Val\`encia, Burjassot, Spain}
}

\maketitle

\begin{abstract}
Sound Event Localization and Detection (SELD) is a problem related to the field of machine listening whose objective is to recognize individual sound events, detect their temporal activity, and estimate their spatial location. Thanks to the emergence of more hard-labeled audio datasets, deep learning techniques have become state-of-the-art solutions. The most common ones are those that implement a convolutional recurrent network (CRNN) having previously transformed the audio signal into multichannel 2D representation. The squeeze-excitation technique can be considered as a convolution enhancement that aims to learn spatial and channel feature maps independently rather than together as standard convolutions do. This is usually achieved by combining some global clustering operators, linear operators and a final calibration between the block input and its learned relationships. This work aims to improve the accuracy results of the baseline CRNN presented in DCASE 2020 Task 3 by adding residual squeeze-excitation (SE) blocks in the convolutional part of the CRNN. The followed procedure involves a grid search of the \textit{ratio} parameter (used in the linear relationships) of the residual SE block, whereas the hyperparameters of the network remain the same as in the baseline. Experiments show that by simply introducing the residual SE blocks, the results obtained are able to improve the baseline considerably.
\end{abstract}

\thispagestyle{firstpage}

\section{Introduction}
\label{sec:intro}

\par Sound Event Localization and Detection (SELD) tries to solve simultaneously two problems related to machine listening: 
tracking the activation of different classes (detection) and localizing spatially those same sound events \cite{Kapka2019, Cao2019, Xue2019}. Detecting the activation of a given sound sound event is known as Sound Event Detection (SED). Unlike other problems such as audio tagging or Acoustic Scene Classification (ASC), solutions for SED have to estimate the onset and offset times of the event. This problem has aroused great interest because of the large number of applications that would benefit from potential solutions. On the other hand, the localization of sound events or direction-of-arrival (DOA) estimation is a problem that has been usually solved with signal processing techniques such as the use of generalized cross-correlations \cite{lucagcc, cobosgcc}. One main limitation of these techniques is that they are not usually able to detect several DOAs at the same time. To mitigate this problem, there are already solutions that implement deep/machine learning techniques to solve this problem independently \cite{vera2018towards, xiao2015learning}. Source positioning information can be very relevant when implementing more robust solutions for SED. Therefore, SELD proposes a joint problem encompassing these two areas, for which a single system capable of performing both detection and location of sound events is desired. In general, to exploit the spatial properties of sound and perform source localization tasks, the system must be fed with signals recorded by a microphone array (multichannel audio input). The popularity of approaches aimed at solving SELD tasks is closely linked to the well-known Detection and Classification of Acoustic Scenes and Events (DCASE) challenge \cite{Adavanne2018_JSTSP, Adavanne2019_DCASE}. In this context, the SELD task first appeared in the DCASE 2019 edition as an extension to the SED problem. 


\begin{figure}[t]
  \centering
  \centerline{\includegraphics[scale=0.54]{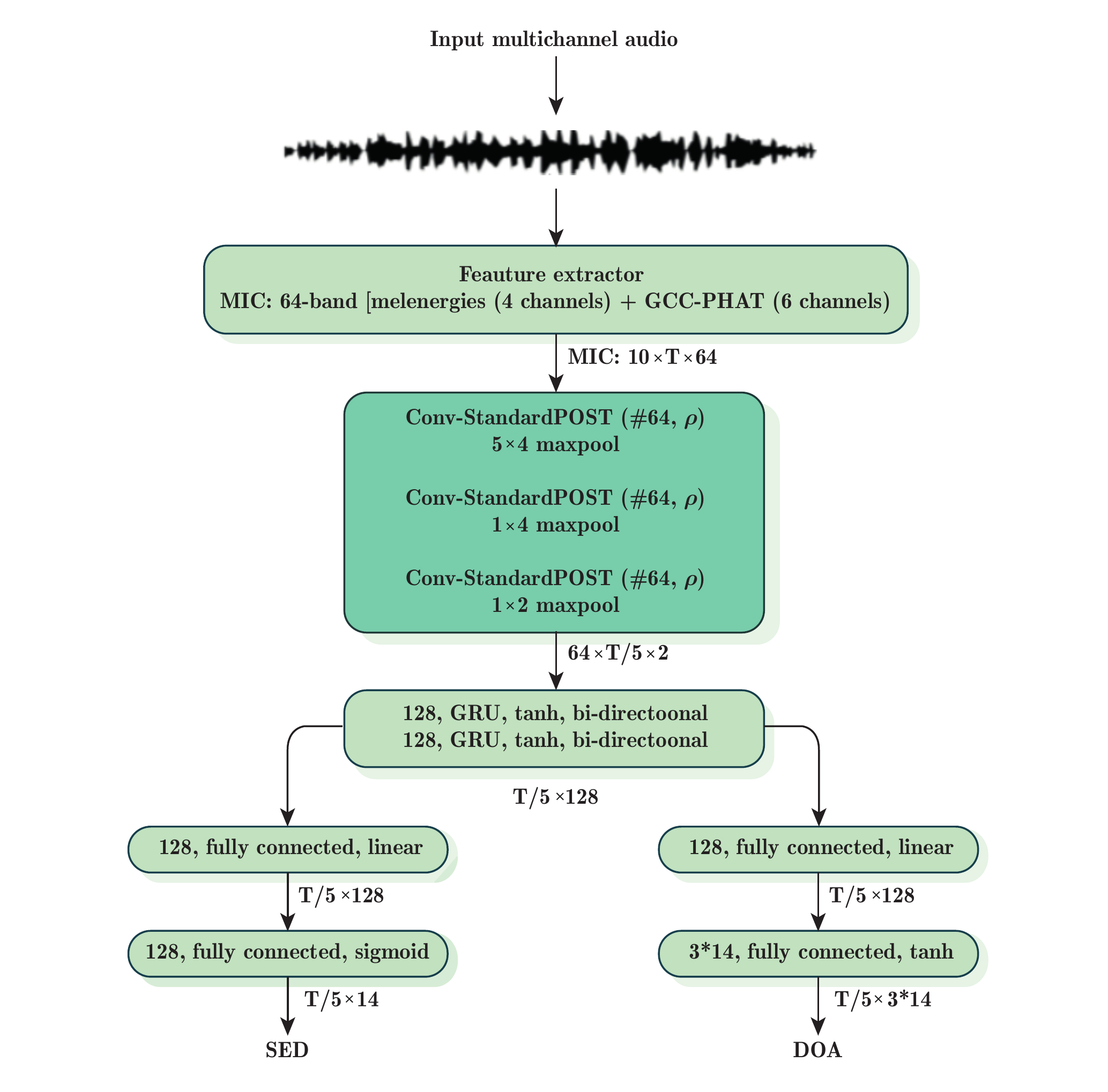}}
  \caption{Modified SELD baseline proposed in this work. The central highlighted block corresponds to the modified convolutional block incorporating SE, with $\rho$ indicating the ratio parameter. The lighter blocks have the same configuration as in the baseline.}
  \label{fig:seld_structure}
\end{figure}

\par Squeeze-excitation (SE) techniques appeared in \cite{hu2018squeeze} as an improvement to the standard convolutional layers. By means of these techniques, the feature maps obtained by the convolutional layers are rescaled. The main idea of squeeze-excitation blocks is to learn spatial and channel-wise feature maps independently instead of jointly as standard CNNs do \cite{roy2018concurrent, naranjoalcazar2020acoustic}. This paper aims to study the improvements that SE techniques can bring to address simultaneously SED and DOA estimation. For this purpose, the convolutional part of the Convolutional Recurrent Neural Network (CRNN) proposed as a baseline in the DCASE 2020 edition is here modified and analyzed, replacing  its convolutional layers by residual SE blocks. As it is the first time that these blocks are introduced in this problem, a grid-search for the hyperparameter \textit{ratio} ($\rho$) of the squeeze block \cite{hu2018squeeze} is performed. The results show that by only introducing this modification, the results of the baseline presented for the SELD task are considerably improved.

\par This paper is organized as follows. Section \ref{sec:method} introduces the baseline network architecture and the modifications applied in this work to incorporate SE techniques. Section~\ref{sec:exp_details} explains the dataset used and the training procedure. Section \ref{sec:results} presents and discusses the results obtained by considering the proposed framework and, finally, Section \ref{sec:conclusion} concludes this article.



\section{Method}\label{sec:method}

\subsection{Baseline system}\label{subsec:baseline}

\par The baseline network considered in this work is known as SELDnet \cite{Adavanne2018_JSTSP}. This network follows a CRNN architecture that uses the confidence of detections (SED) to estimate the DOA of each of the classes. The SED is displayed as a multi-label classification and the DOA as a multi-output regression.

\par The input to the network is defined as a 10-channel time-frequency representation of the input sound. Audio signals have been recorded with a 4-mic array. Therefore, four of the input channels are log-Mel spectrogram representations of the mic signals, whereas the other six inputs are time-frequency representations of the Generalized Cross-Correlations (GCCs) between the mic signals. A more in-depth description of the audio dataset of this task is given in Section~\ref{subsec:dataset}. The dimension of each channel is $T \times F$, where $T$ corresponds to the number of time frames and $F$ to the number of frequency bins. In this case, $F$ is set to 64 and $T$ corresponds to 300 temporal bins. In order to reach this temporal resolution, the sampling frequency is set to 24000 Hz, the window size is 40 ms and the overlap is 50\%. Mel-scale filters are designed to cover the full frequency range (from 0 Hz to 12000 Hz). These hyperparameters are the same as those presented in the baseline.


\begin{figure}[t]
  \centering
  \centerline{\includegraphics[scale=0.8]{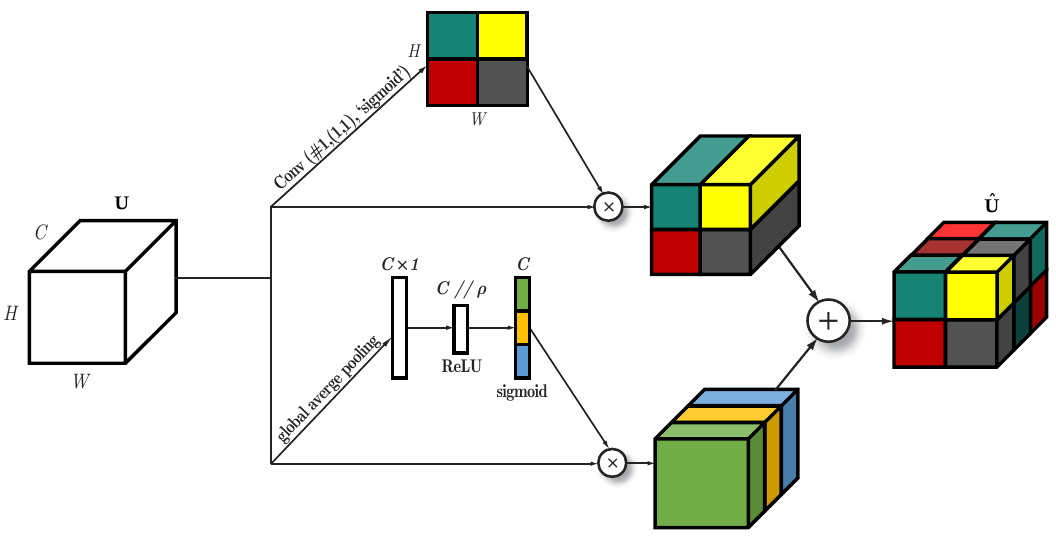}}
  \caption{Diagram of an scSE block, composed by an spatial Squeeze-Excitation (sSE) module (top branch) and a channel Squeeze-Excitation (cSE) module (lower branch). \cite{naranjoalcazar2020acoustic, roy2018concurrent}} 
  \label{fig:scSE}
\end{figure}

\subsection{Squeeze-Excitation}\label{subsec:SE}

\par Most machine listening frameworks rely on the ability of the CNN to extract meaningful features. Either in a VGG-style \cite{simonyan2014very} or ResNet form \cite{he2016deep}, many networks share common similarities. It is well known that overall performance improvements come many times from other aspects, such as the use of data augmentation techniques (pitch shifting \cite{Chang2019}, speed perturbation \cite{Xue2019} or mixup \cite{Jee2019} among others) or the ensemble of many independent models \cite{Xue2019, Nguyen2019}. 

\par In \cite{naranjoalcazar2020acoustic}, the authors presented an analysis of different residual SE blocks suggested in \cite{hu2018squeeze} and proposed two alternative blocks making use of the \textit{Concurrent Spatial and Channel Squeeze and Channel Excitation} (scSE) configuration introduced in \cite{roy2018concurrent}. As showin in Figure~\ref{fig:scSE}, the block is parameterized by a single parameter ($\rho$), corresponding to the ratio 
specified in the lower branch of the scSE block. This parameter establishes a reslationship between the number of channels and the number of units in the following dense layer. In this work, a grid-search is performed to analyze a proper $\rho$ value fitting this problem. The analysis is carried out without any data augmentation technique during training. 

\begin{figure}[t]
  \centering
  \centerline{\includegraphics[scale=0.6]{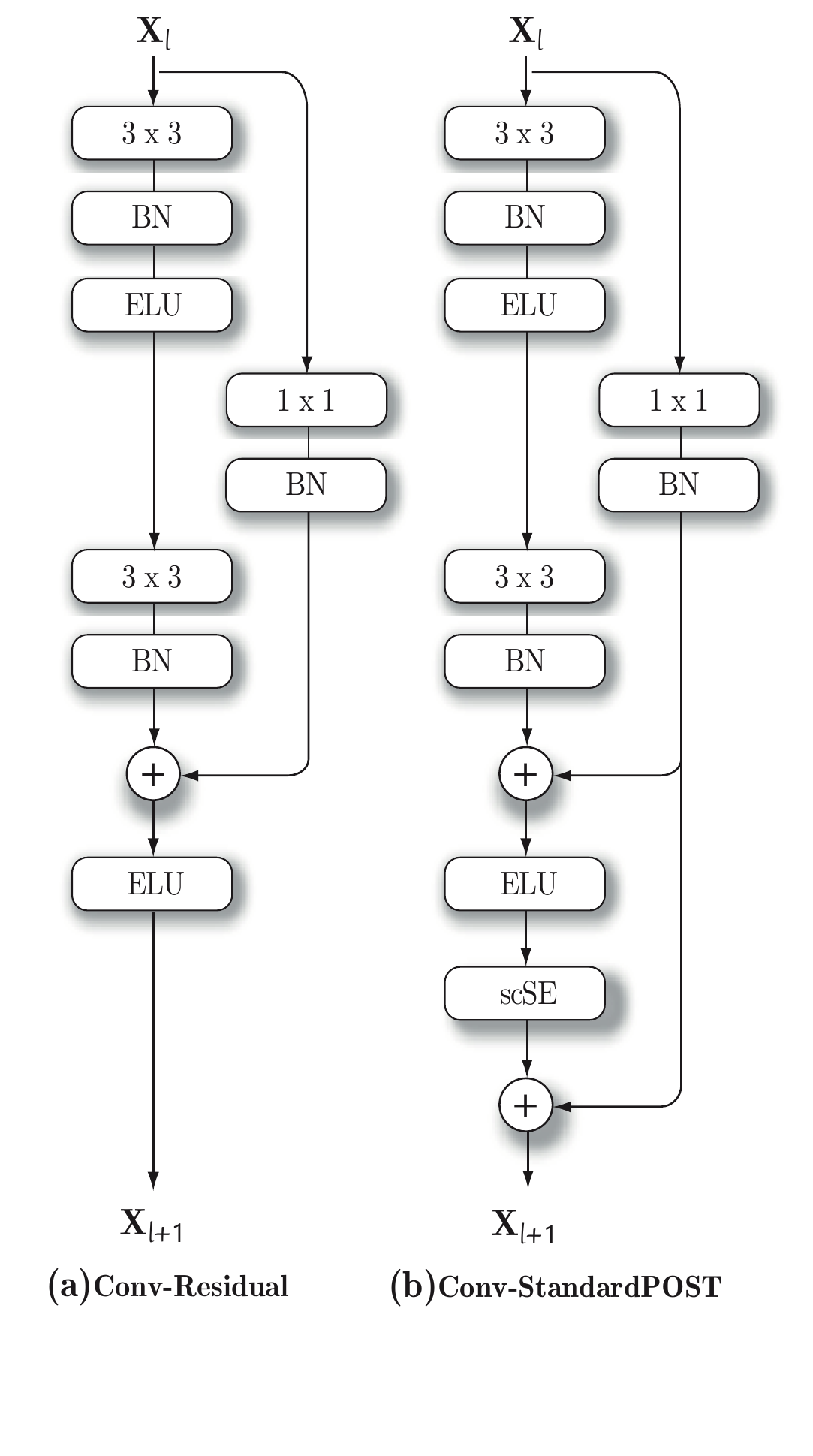}}
  \caption{Residual blocks analyzed in this paper. The internal layers correspond to Batch Normalization (BN), squeeze-excitation (scSE) and convolutional layers (denoted with their kernel size).}
  \label{fig:seld_residual}
\end{figure}

In \cite{naranjoalcazar2020acoustic}, the best results for an acoustic scene classification task were obtained with one of the proposed scSE blocks, denoted as \textit{Conv-StandardPOST}, and it is the one used here to explore the benefits of SE within an SELD context. In addition, to analyze better the contribution of SE, the network was also trained with a conventional residual configuration, denoted as \textit{Conv-Residual} in \cite{naranjoalcazar2020acoustic}. The structures of the two blocks above are depicted in Figure~\ref{fig:seld_residual}(b)). Further details on these and other alternative SE blocks are discussed in \cite{naranjoalcazar2020acoustic}. The code used for this experimentation can be found in this link\footnote{https://github.com/Joferesp/DCASE2020-Task3}.

\section{Experimental details}\label{sec:exp_details}

\subsection{Dataset}\label{subsec:dataset}

\par The dataset used for this work is called TAU-NIGENS Spatial Sound Events 2020 \cite{Adavanne2019_DCASE}. It should be observed that each scene is delivered in two different formats: First-Order Ambisonics (FOA) and the four signals captured by a tetrahedral microphone array (MIC). In both formats (MIC or FOA), each sound event in the scene is associated with a DOA to the recording point, including as well timestamps corresponding to its onset and offset times. The number of classes to be detected are 14. Some of these classes are: piano, male speech, female speech, barking dong, among others. Note that sounds belonging to the aforementioned classes are easily found in domestic environments, encouraging solutions that could improve real-world applications such as home assistants \cite{sigtia2016automatic}. For this work, only MIC recording format has been used.




\par  Concerning the usage of the samples, in the development phase, three folders are used for training, one for validation and one for testing. In this stage, the ground truth of all the samples is available. However, in the evaluation stage, 4 folders are used for training, 1 for validation and 2 for testing. In this case, the ground truth of the test samples is not available, therefore, the results of the validation folder will be reported in this stage. The distribution can be seen in Table~\ref{tab:kfold}. Each folder consists of 100 audio samples of 1 minute duration. Thus, in the development phase, 600 labelled samples are available. In the evaluation stage, 200 more samples are added but without the corresponding labels.


\begin{table}[]
\centering
\caption{Distribution of the folders in the two stages. Each folder contains 100 samples.}
\begin{tabular}{cccc}
\toprule
Stage       & training & validation & test \\ \toprule
Development & 3-6      & 2          & 1    \\ \midrule
Evaluation  & 2-6      & 1          & 7-8 \\ \bottomrule
\end{tabular}
\label{tab:kfold}
\end{table}

\subsection{Training procedure}\label{subsec:training}

\par The training process is the same as that proposed in the baseline \cite{Adavanne2018_JSTSP, politis2020dataset}. The optimiser used for training was Adam with learning rate of $0.001$. The training consisted of a total of 50 epochs. No hyperparameter, such as learning rate, the decay weight, number of epochs, etc., was modified; in this way, the variations in the results can only be attributed to the proposed modifications explained in Section~\ref{subsec:SE}.

\section{Results}\label{sec:results}

\par In order to carefully study the effect of SE residual blocks, it was decided to carry out a grid search over the ratio parameter. We emphasize here that the network is made up of 3 blocks of 64 filters. The ratio ($\rho$) is the same for all blocks, as it can be seen in Figure~\ref{fig:seld_structure}. 

\subsection{Metrics}\label{subsec:metrics}

\par For evaluation, we use the same metrics as the ones used in the SELD task of the DCASE 2020 edition, which evaluate jointly the localization and detection performance \cite{Mesaros_2019_WASPAA}. A prediction will be considered correct if there is a match between the predicted and the true class of the event and if the difference between the predicted and true DOA angle is below 20º. Thus, the conventional detection metrics are now location-dependent. The traditional error rate, ER\textsubscript{20º}, and F-score, F\textsubscript{20º}, metrics are used for the evaluation of the detection performance, where the subscript 20º means that only the events where the error in the predicted angle in below 20º are considered as positives. On the other hand, for evaluating localization performance, the metrics used are the localization error (LE\textsubscript{CD}), expressing average angular distance between predictions and references of the same class, and the localization recall metric (LR\textsubscript{CD}), which expresses the true positive rate of how many of these location predictions were detected in a class, of the total occurrences of such class. Metrics used in the 2019 edition are also reported \cite{Mesaros2016_MDPI} for completeness. As the results of the development and evaluation phases differ, the section has been divided into two separate subsections.

\subsection{Development stage}\label{subsec:dev}

The results for the data provided in the development stage are in Tables \ref{tab:results_2019} and \ref{tab:results_2020}. The first column indicates the evaluated system, where \textit{Conv-Residual} corresponds to the residual block shown in Figure~\ref{fig:seld_residual}(a) and $\rho$ indicates the particular ratio value used in the \textit{Conv-StandardPOST} configuration of Figure~\ref{fig:seld_residual}(b).

\begin{table}[H]
\centering
\caption{Development results using DCASE2019 metrics \texttt{(`dev')}.}
\scalebox{0.87}{
\begin{tabular}{ccccc}
\toprule
\textbf{framework} & \textbf{ER}   & \textbf{F (\%)}    & \textbf{LE (º)}   & \textbf{LR (\%)} \\
\toprule
\textit{baseline} & \textit{0.56} & \textit{59.2} & \textit{22.6} & \textit{66.8} \\ \midrule
\textit{Conv-Residual}     & 0.50 & \textbf{65.2} & 19.0 & 68.5  \\ \midrule
$\rho$ = 1     & 0.51 & 63.7 & 20.5   & \textbf{69.1}  \\ \midrule
$\rho$ = 2     & 0.52 & 62.2 & 19.4 & 68.1  \\ \midrule
$\rho$ = 4     & \textbf{0.49} & 65.1 & 20.2 & 68.1  \\ \midrule
$\rho$ = 8     & 0.51 & 64.0 & 19.4 & 67.4  \\ \midrule
$\rho$ = 16    & 0.52 & 63.0 & \textbf{18.6} & 68.0  \\ \bottomrule
\end{tabular}}
\label{tab:results_2019}
\end{table}

\begin{table}[H]
\centering
\caption{Development results using DCASE2020 metrics \texttt{(`dev')}.}
\scalebox{0.87}{
\begin{tabular}{ccccc}
\toprule
\textbf{framework} & \textbf{ER\textsubscript{20º}}   & \textbf{F\textsubscript{20º} (\%)}    & \textbf{LE\textsubscript{CD} (º)}   & \textbf{LR\textsubscript{CD} (\%)} \\
\toprule
\textit{baseline} & \textit{0.78} & \textit{31.4} & \textit{27.3} & \textit{59.0} \\ \midrule
\textit{Conv-Residual}     & \textbf{0.68} & \textbf{42.3} & \textbf{22.5} & \textbf{65.1}  \\ \midrule
$\rho$ = 1     & 0.70 & 39.2 & 23.5 & 63.6  \\ \midrule
$\rho$ = 2     & 0.69 & 40.4 & 23.2 & 62.1  \\ \midrule
$\rho$ = 4     & \textbf{0.68} & 40.9 & 23.3 & 65.0  \\ \midrule
$\rho$ = 8     & 0.69 & 40.8 & 23.5 & 63.8  \\ \midrule
$\rho$ = 16    & 0.69 & 40.7 & 23.3 & 62.8  \\ \bottomrule
\end{tabular}}
\label{tab:results_2020}
\end{table}

\par As it can be observed in Table~\ref{tab:results_2019}, all the residual configurations outperform significantly the baseline. Residual learning allows obtaining more accurate systems by adding only a shortcut. Moreover, the results suggest that the concerned SE improves the results in all metrics except in $\mathbf{F}$. However, the best value for $\rho$ is different depending on the considered metric. In contrast, when considering the results of Table~\ref{tab:results_2020}, the results with \emph{Conv-Residual} are now the best. This might suggest that when the metrics are restricted to consider only correctly localized events, SE techniques may perform slightly worse. However, as described in the next section, this may not be the case, as the use of more training examples seems to favor SE blocks.


\subsection{Evaluation stage}\label{subsec:eval}

To complete the study, we present the results over the same conditions considered in the DCASE2020 evaluation stage. In this case, the ground-truth data for the test folds is not available and the results are directly provided by the organizers for a maximum of 4 systems. The four evaluated systems correspond to those that obtained the best results on the validation folder in the evaluation stage (folder 1).


\begin{table}[H]
\centering
\caption{Evaluation results using DCASE2020 metrics \texttt{(`eval')}.}
\scalebox{0.87}{
\begin{tabular}{ccccc}
\toprule
\textbf{framework} & \textbf{ER\textsubscript{20º}}   & \textbf{F\textsubscript{20º} (\%)}    & \textbf{LE\textsubscript{CD} (º)}   & \textbf{LR\textsubscript{CD} (\%)} \\
\toprule
\textit{baseline} & \textit{0.69} & \textit{41.3} & \textit{23.1} & \textit{62.4} \\ \midrule
\textit{Conv-Residual}     & \textbf{0.61} & 48.3  & \textbf{19.2} & 65.9 \\ \midrule
$\rho$ = 1    & \textbf{0.61} & \textbf{49.1} & 19.5 & \textbf{67.1}  \\ \midrule
$\rho$ = 8    & 0.64 & 46.7 & 20.0 & 64.5  \\ \midrule
$\rho$ = 16   & 0.63 & 47.3 & 19.5 & 65.5   \\ \bottomrule
\end{tabular}}
\label{tab:results_2020_eval}
\end{table}

As observed in Table~\ref{tab:results_2020_eval}, in evaluation step, the SE alternative with $\rho = 1$ slightly outperforms the \textit{Conv-Residual} with the 2020 metrics except for the localisation error. This may confirm that SE techniques contribute to more accurate results in SELD tasks but, at the same time, they may require more data to learn relationships that can generalize better in the test stage. With the considered data partition, the results suggest that the implementation of the \textit{Conv-StandardPOST} block with $\rho = 1$ shows the best trade-off between the SED and DOA tasks. Unfortunately, 2019 baseline metrics are not available at the evaluation stage.

\section{Conclusion}\label{sec:conclusion}

This paper analyzed the improvements that residual learning and squeeze excitation (SE) techniques can bring to the field of sound event localization and detection. To this end, we applied slight modifications to a baseline by replacing their conventional blocks with modified blocks implementing residual learning and SE. To simplify comparisons, we did not include data augmentation or model ensembling techniques during training. The results confirm that both residual learning and SE affect positively the performance of the resulting models, although the use of SE techniques may require more training data for obtaining better performance.


\section*{Acknowledgements}

\par The participation of Javier Naranjo-Alcazar and Dr. Pedro Zuccarello was partially supported by fellowships DIN2018-009982 and PTQ-17-09106, respectively, from the Spanish Ministry of Science, Innovation and Universities. The work of Maximo Cobos and Francesc J. Ferri was partially supported by the Spanish Ministry of Science, Innovation, and Universities and the European Regional Development Fund (ERDF) under Grant RTI2018-097045-B-C21, and partially by Generalitat Valenciana under Grants AICO/2020/154 and AEST/2020/012.


\bibliographystyle{IEEEtran}
\bibliography{refsursi}

\end{document}